\documentstyle[preprint,aps]{revtex}

\begin{document}

\draft

\title{CLOSE PACKING OF ATOMS, GEOMETRIC FRUSTRATIONS AND \\
THE FORMATION OF HETEROGENEOUS STATES IN CRYSTALS}
\author{Yu.~N.~Gornostyrev$^{1}$, M.I.Katsnelson$^{1}$, A.V.Trefilov$^{2}$}

\address{
\(^1\) Institute of Metal Physics, Ekaterinburg, Russia. \\
\(^2\) Kurchatov Institute, Moscow, Russia.}

\maketitle


\begin{abstract}
To describe structural peculiarities in inhomogeneous media caused by the
tendency to the close packing of atoms a formalism based on the using of the
Riemann geometry methods (which were successfully applied lately to the
description of structures of quasicrystals and glasses) is developed. Basing
on this formalism we find in particular the criterion of stability of
precipitates of the Frank-Kasper phases in metallic systems. The nature of
the ''rhenium effect'' in W-Re alloys is discussed.
\end{abstract}

\vskip 0.5cm

Recently the interest grows to the significance of crystallogeometrical
factors, first of all connected with the close packing of atoms, for the
structure of condensed matter\cite{1}. There is a whole class of metallic
systems which structure is apparently determined by the tendency to the
close packing of ionic spheres with different radii including quasicrystals,
metallic glasses \cite{1}, small metallic clusters \cite{2} and Frank-Kasper
(FK) phases \cite{3}. It is worthwhile to stress that the considerations of
the closest local packing may contradict with the existence of long-range
crystal order. It distinguishes essentially the real three-dimensional case
from the two-dimensional one where (for the case of spheres of equal radii)
the triangle lattice provides the closest packing both for the whole space
and its every part i.e. both globally and locally \cite{4}. Classical
problem about the distribution of spheres in three-dimensional space to
obtain the closest packing has not yet solved rigorously (it is the part of
the 18th Gilbert problem). However, it is known that spheres may be packed
locally with higher density than in fcc, hcp and other close packed lattices
\cite{4}. Therefore the situation of geometric frustrations arises when the
optimum structure from the point of view of local surrounding cannot be
optimum globally. The concept of frustrations appeared to be useful in
particular for the consideration of the structure of disordered systems \cite
{1,5}.

The cause of the existence of the geometric frustrations is the fact that
the {\it Euclidean } space cannot be occupied completely by regular
tetrahedra (the regular tetrahedron is the structure unit providing the
closest packing of four atom groups). An elegant technique has been proposed
in Ref.6 to construct the structures of FK phases when the latter are
obtained from the closest packing of tetrahedrons in {\it Riemannian} space
by the introducing of the net of structure disclinations (SD) decurved the
space and then the filling of the Euclidean space by the structure units
arisen.Here we will use a similar approach to analyze the problem of
formation of heterogeneous state (HS) in metallic alloys. Such state is the
subject of great interest of researchers in the field of material science
lately. From the experimental point of view it is distinguished by the
peculiar diffuse X-ray scattering corresponding in the real space to the
presence of relatively small clusters containing only hundreds or even tens
of atoms \cite{7}. The examples are the so called athermic $\omega $-phase
in some Ti- and Zr-based alloys \cite{8} and in Cr$_{1-x}$Al$_x$ \cite{9}, $%
\sigma $-phase in Fe$_{1-x}$Cr$_x$ \cite{10}, precipitates of W$_3$Re phase
with A15 structure in W$_{1-x}$Re$_x$ \cite{11}. Recent investigations of Ti$%
_{1-x}$Fe$_x$ alloys by Moessbauer effect \cite{12} have shown the presence
of geometric frustrations i.e. the difference of the type of short-range
order in local surrounding of Fe nuclei from the long-range order in a
crystal as a whole. The hypothesis has been proposed in Ref. 13 about the
existence of small icosahedral clusters in this alloys.

Generally speaking, the structure of crystal phases is determined by the
whole number of factors. Among them, apart from the considerations of close
packing, the spatial orientation of covalent bonds, peculiarities of
electronic structure near the Fermi surface etc are the most frequently
discussed. Therefore it is naturally to expect that the nature of HS may be
different for different systems. In particular Krivoglaz  \cite{7} stressed
the role of peculiar features of the shape of the Fermi surface for the
formation of HS in Ti- and Zr-based alloys.This mechanism is however hardly
to be universal since in the whole number of systems with HS the electron
mean free path is small and therefore the alloy smearing of the Fermi
surface has to be essential.

The alternative mechanism is proposed in the present work basing on the
considerations of the closest packing of atoms for small atomic groups and
geometric frustrations connected with them which may lead to the formation
of HS with characteristic scale of order of few interatomic distances.The
model proposed is rather rough and pretends to describe only some main
features of the phenomenon. Nevertheless it stress, to our opinion, a
peculiar role of the factor which is important for many real systems but was
not taken into account in previous works.

Consider the precipitates of A15 phase in bcc host (e.g. of W$_3$Re phase in
the solid solution W$_{1-x}$Re$_x$ [11]) as an example of heterogeneous
state of the type under consideration because A15 structure is the simplest
and most well-known representative of FK phases \cite{1}. It contains eight
atoms per cell, two of them having icosahedral coordination and six of them
being surrounded by polyhedra with 14 vertices \cite{6}. The latter may be
obtained from the regular icosahedron by the introducing of the edge
disclination with the axis passing its center and the Frank vector -$2\pi /5$%
. Using such a procedure it was shown in Ref.6 that the net of SD for A15
structure consists from three mutually orthogonal sets of equidistant
parallel disclinations with the period being equal to the lattice constant $a
$. Other FK phases differs from A15 structure only by the geometry of SD net.

To solve the question about the stability of HS one needs to generalize the
approach \cite{6}. For this aim we use the basic relation between the
curvature tensor $R_{ijkl}$ and the disclination density tensor $\theta
_{ij}^{(c)}$ \cite{14}
$$
R_{ijkl}=-\epsilon _{ijp}\epsilon _{klq}\theta _{pq}^{(c)},\eqno(1)
$$
where $\epsilon _{ijp}$ is the unit antisymmetric tensor. In light of Eq.
(1) the space curvature providing the ideal tetrahedral packing in the
approach \cite{1,6} is created by the introducing of the partial edge
disclinations with Frank vector $\Omega =7^{\circ }20^{\prime }$ which
compensate the deficiency of dihedral angles at the packing of five
tetrahedra around the common edge \cite{15}. To construct FK phases it is
necessary to ``decurve'' the space by the introducing of SD with the average
density $\overline{\theta }_{ij}^{SD}=-\theta _{ij}^{(c)}$ where the line
means the average over the volume containing large enough number of cells.
Thus, in contrast with the approach of Ref.6 where SD were introduced in
{\it curved} space we introduce {\it two} sets of disclinations in {\it %
Euclidean} space with the mean total density being equal to zero. At the
same time local variation of the curvative of the lattice (i.e. the
deviation of {\it local} disclination density from zero) is possible and
connected with the elastic distortions of the bonds. The corresponding
stresses are similar to that which have discussed in Ref.5 for glasses.

Apart from the energy of elastic distortions the electronic (``chemical'')
contribution to the energy describing the tendency to the closest packing of
atoms also exist. If in some place the density of SD differs from the mean
density in the FK phase and therefore according to (1) the lattice appears
to be locally curved the local values of the packing density will be higher
or lower than in the FK phase.

In the framework of the approach considered the energy of the FK phase
counted from the energy of the ''host'' phase with $\theta =0$ may be
written as
$$
E=\int d{\bf r}f\left[ \theta _{ij}({\bf r})\right] +E_{el},\eqno(2)
$$
where $f$ is the density of ``chemical'' energy depending on the degree of
the atomic packing which is connected with the disclination density tensor $%
\theta _{ij}$, $E_{el}$ is the elastic energy. As usual \cite{16} the latter
may be represented in linear elasticity theory in the following form
$$
E_{el}=\frac 12\int \int d{\bf r}d{\bf r}^{\prime }\eta ^{\nu \rho }({\bf r}%
)H_{\nu \rho \kappa \tau }({\bf r}-{\bf r}^{\prime })\eta ^{\kappa \tau }(%
{\bf r}^{\prime }),\eqno(3)
$$
where $H_{\nu \rho \kappa \tau }$ is the Green tensor for internal stresses,
$\eta _{\nu \rho }$ is the incompatibility tensor describing the density of
sources of internal stresses. According to \cite{14} it is defined as
$$
\eta ^{\nu \rho }=-\frac 12(\epsilon _{\rho pq}\alpha _{\nu q,p}+\theta
_{\nu \rho }+\epsilon _{\nu pq}\alpha _{\rho q,p}+\theta _{\rho \nu }),%
\eqno(4)
$$
where $\alpha _{\nu \rho }$ is the disclination density tensor. Separating
the singular part of the Green tensor we may represent the energy  $E_{el}$
as the sum  $E_{el}=E_{el}^{(0)}+E_{el}^{(1)}$ of the energy of distortions
in the disclination cores $E_{el}^{(0)}$ and the energy of elastic
deformations outside the cores $E_{el}^{(1)}$.

When dislocations are absent one has $\eta _{\nu \rho }=-1/2(\theta _{\nu
\rho }+\theta _{\rho \nu })$. For single disclination with the Frank vector
(0,0,$\Omega $) the tensor $\eta _{\nu \rho }=-\theta _{\nu \rho }$ has the
only non zero component \cite{14}
$$
\eta ^{33}=-\Omega \delta ({\bf \rho }),\eqno(5)
$$
where ${\rho }=(x,y)$. In the case of the only set of edge disclinations
considered here the tensor $\theta _{ij}$ according to (5) may be
characterized by the only scalar parameter $\theta =Tr\theta _{ij}$ which
equals to the mean value of the Frank vector in a given point. It is
proportional to the scalar curvature $R_{ijij}$ (see (1)). For the given
geometry the energy of disclination cores (per unit length) may be
represented as
$$
\frac{E_{el}^{(0)}}L=e_0\int d^2\rho \theta ^2(\rho )
$$
where  $e_0$ is the energy of the core of the disclination with unit Frank
vector determining by the relation $e_0\delta (\rho )=\frac 12H_{3333}(0)$.
Note that for the elastic continuum $e_0=0$ and it is necessary to use the
quasicontinuum model \cite{16} to describe the core energy correctly.

It is a common practice in the material science to treat inhomogeneous state
mainly as the multiphase state with the coexistence of regions with
different crystal structures each of them may be stable in principle in the
whole space. In this case the inhomogeneity may be described in terms of
elastic distortions due to the conjugation of crystal lattices on the
boundary of the precipitate and the host. To calculate the distortions the
Eshelby model \cite{Eshelby} is usually used, their energy appearing to be
proportional to the volume of the precipitate. It is the consequence of the
fact that in the Eshelby model the deformation $\varepsilon _{ij}^{(0)}$
connected with the phase transition is constant inside the precipitate.
Then, according to Eshelby,  the deformation $\varepsilon _{ij}^{(0)}$ may be
simulated by a system of disclination loops on the boundary of the
precipitate. Since the energy of every loop is proportional to its radius $R$
(equal to precipitate radius) and the number of loops is proportional to $R^2$ 
their total energy turns out to be proportional to $R^3$. In our model the 
precipitate has the type of short-range order (e.g. icosahedral) which cannot 
correspond to long-range order for any bulk crystal phase. Therefore in contrast 
with the Eshelby model the deformation appears to be inhomogeneous not only outside
the precipitate but also inside it. As the result it will be shown below
that the elastic energy of the precipitate grows with the increase of  $R$
faster than  $R^3$ and therefore the stabilization of HS becomes possible.

The real form of the function $f(\theta )$ in (2) is unknown. In accordance
with our choice of the zero point for the energy we have for host phase $%
(\theta =0)$ $f(0)=0$ and the minimum of $f(\theta )$ lies at the value $%
\theta =\theta _0$ corresponding to the closest tetrahedral packing in the
Riemannian space \cite{2}. We suppose for the simplicity
$$
f=\alpha (\theta -2\theta _0)\theta ,\ \,(0<\theta <\theta _0),\eqno(6)
$$
where the parameter $\alpha >0$ depends on the explicit form of the
interatomic interactions. Generally speaking, the function  $f$ may contain
also terms proportional to  $(\nabla \theta )^2$, but in the framework of
the variational approach used below (see (8)) they do not influence on the
results.

To demonstrate the possibility of the formation of heterogeneous state we
use the direct variational approach and construct explicitly the
distribution $\theta (r)\neq const$ leading to the lower value of the total
energy than the homogeneous state. The problem may be solved in the simplest
way for the cylindrical precipitate of the close packed phase. Let $0z$ be
the axis and $R$ the radius of the precipitate, the axis of all the
disclinations are parallel to $0z$ and $\theta =\theta (\rho )$ where $\rho =%
\sqrt{x^2+y^2}$.

Since it is obvious that the discontinuities on the boundary between the
host and precipitate are energetically unfavorable we restrict ourselves by
the case of their coherent conjugation when the discontinuities are absent.
In this case the disclinations cannot abrupt at the boundary and should form
loops with external segments in the host. In contrast with the homogeneous
A15 phase the ''polarized'' distribution of disclinations is typical for our
case namely the uniform distribution of positive partial disclinations inside
the precipitate and the ''cloud'' of negative disclinations outside it. From
topological considerations (the vanishing of the total Frank vector) one has
$$
\int d^2\rho \theta (\rho )=0\eqno(7)
$$
To evaluate the minimum of the energy (2) we use the trial function
$$
\theta (\rho )=\left\{
\begin{array}{lc}
{\theta _1,\text{ \ \ \ \ \ \ \ \ \ }\rho <R} &  \\
{\frac{-\theta _1R^2}{2R\Delta +\Delta ^2},\text{ \ \ }{R<\rho <R+\Delta }}
&  \\
{0,\text{ \ \ \ \ \ \ \ \ \ \ }{\rho >R+\Delta }} &
\end{array}
\right. \eqno(8)
$$
answering the requirement (7) automatically. The value $\Delta $ describes
the thickness of the ''cloud'' and should be found from the solution of the
variational problem. In the model under consideration the energy of elastic
distortions concentrated in the disclination cores in the cloud would play
the role of the surface energy.  However, due to the condition (7) the total
number of disclinations in the cloud is equal to that in the precipitate
itself and is proportional to its volume (in 2D case  $\sim R^2$). Therefore
in contrast with the Eshelby model it is impossible to separate explicitly
the bulk and surface parts from the precipitate energy.

We use for $H_{3333}(\rho -\rho ^{\prime })=H(\rho -\rho ^{\prime })$ the
known expression \cite{16}
$$
H(r)=-\frac{2\mu }{1-\nu }\frac{\rho ^2}{8\pi }\left[ 1-ln\frac \rho
{R_c}\right] +C\eqno(9)
$$
corresponding to the continual approximation in the elastically isotropic
model where $\mu $ is the shear modulus, $\nu $ is the Poisson ratio, $R_c$
is the cutoff radius of order of the size of a crystal, $C$ is the constant
which does not contribute to the energy $E_{el}^{(1)}$ for a given geometry
(total Frank vector equals zero). To describe the energy of disclination
cores $E_{el}^{(0)}$correctly one should use the quasicontinuum
approximation  \cite{16}
$$
C=-\frac \mu {2\pi (1-\nu )k_d^2}\left[ 1+2\gamma \ln k_dR_c\right]
$$
where $k_d$ is the Debye wave vector,  $\gamma $ is the parameter depending
on the explicit form of phonon dispersion curves in the model.

Then the elastic energy (3) per unit length of the precipitate does not
depend on the cutoff radius $R_c$  and has the form
$$
\frac{E_{el}^{(1)}}L=\theta _1^2\frac{\mu a^4}{1-\nu }\widetilde{\psi }%
(R,\Delta )\eqno(10)
$$
where
$$
\widetilde{\psi }(R,\Delta )=\frac{R^4(\Delta +R)^2}{192a^4\Delta ^2\left(
\Delta +2R\right) ^2}
$$
$$
\left[ {\Delta }\,(5\,{{\Delta }^3}+20\,{{\Delta }^2}\,R+26\,{\Delta }\,{R^2}%
+12\,{R^3})\right.
$$
$$
\left. -12\,{R^2}\,{(\Delta +R)}^2\,\ln \left( \frac{\Delta +R}R\right)
\right]
$$
The minimizing of the function $\widetilde{\psi }(R,\Delta )$ with respect
to $\Delta $ gives the dependence of the cloud thickness $\Delta =\Delta (R)$%
. As a result the total energy may be represented in the following form
$$
\frac EL=\pi R^2\alpha p(\theta _1-2\widetilde{\theta }_0)\theta _1+\theta
_1^2\frac{\mu a^4}{1-\nu }\psi (R)\eqno(11)
$$
where $\psi (R)=\widetilde{\psi }(R,\Delta (R))$, $\widetilde{\theta }%
_0=\theta _0/p,p=1+2e_0/\alpha $. The minimum of energy (11) corresponds to
the disclination density
$$
\theta _1=\frac{\widetilde \theta _0}{1+\psi (R)/\kappa R^2},\eqno(12)
$$
where the precipitate radius $R$ is found from the equation
$$
R\frac{d\psi (R)}{dR}=2\left( 2\psi (R)+\kappa {R^2}\right) \eqno(13)
$$
Here
$$
\kappa =\frac{\delta f(1-\nu )\pi p}{\mu a^4\theta _0^2}\eqno(14)
$$
and the value $\delta f=\alpha \theta _0^2$ describes the chemical energy
gain (per unit volume) at the transition to the ideal tetrahedral packing.

As it follows from Eq. (13) the precipitate radius $R$ depends on the
parameter $\kappa $ characterizing the ratio of the chemical contribution to
the elastic one. According to (14) its value varies from 0.1 to 1 at $\delta
f\approx 0.01-0.1$ eV per atom. Direct calculations show that in all these
limits of the variation of $\kappa $ the function $\psi (R)$ is almost
constant at $R<2a$ and increases sharply at $R>3a$. On that ground, the
equilibrium value of $R$ grows slowly with $\kappa $ increase (see Fig.1).
At $\kappa <0.1$ the continual approach under consideration does not hold
because $R<a$. In this range of values of $\kappa $ the disclination density
$\theta _1$ determined by Eq. (12) is less than the minimal possible value
corresponding one disclination with Frank vector $7^{\circ }20^{\prime }$
per precipitate.

The spatial distribution of the dilatation obtained by numerical integration
is shown in Fig.2. One may see that the internal part of the precipitate is
expanded and the dilatation $\varepsilon _{ll}$ vanishes sharply at the
boundary. The mean value in the precipitate  $<\varepsilon _{ll}>$ increares
approximately linear with the precipitate radius. This demonstrates the
qualitative difference between the model under consideration and the Eshelby
model. Since the average deformation increases with $R$ increasing the
elastic energy will grow faster than  $R^2$ in  2D case ( or than  $R^3$ for
spherical precipitate).

Putting the equilibrium values of $\theta _1,R$ in Eq.(11) one obtains the
expressions for the energy change at the formation of the precipitate
$$
\frac{ E}L=-\delta f\pi R^2\frac{\theta _1}{\theta _0}=-\frac{\theta
_0\theta _1^2}{\theta _0-\theta _1}\frac{\mu a^4\psi (R)}{1-\nu }<0\eqno(15)
$$
Thus in the model under consideration the precipitates are stable and the
heterogeneous state has the lower energy than the bcc lattice. Accordingly
to the dependence $\theta _1$ on $R$ its value (12) , generally speaking, is
not coincide with one of the values corresponding homogeneous FK phases.

Thus we show that for certain ratio between ''chemical'' and elastic
properties of the media ($\kappa \geq 0.1)$ the heterogeneous state may be
energetically favorable owing to the crystallogeometrical factors (tendency
to the closest local packing). It is worthwhile to stress again the most
important features of this state in the framework of the model under
consideration (i) sizes of the precipitates are of order of few lattice
constants (ii) the structure of precipitates is distorted in comparison with
the structures of FK phases in an infinite crystal (iii) precipitates are
surrounded by the cloud of the distorted host lattice with the thickness of
order of the sizes of the precipitates. The existence of this cloud has an
obvious sense: since it is impossible to fill the whole space by the regular
tetrahedra an attempt to realize such filling locally in some part of the
space results inevitably to the appearance of voids at the boundary. The
introducing of the external segments of disclination loops lead to the
elimination of these voids and coherent conjugation of the precipitate with
the host surrounded.

In the specific case when the host phase is bcc a sharp increase of the
solubility of light interstitial impurities is one of the characteristic
features of  the mechanism considered of the formation of HS. It is
connected with the presence of large tetrahedral voids in close packed
structures and their absence in bcc lattice. This feature together with the
change of the type of short-range order and the appearance of elastic
distortions can help to identify this kind of HS experimantally.

The results obtained here may explain some important features of the so
called ''rhenium effect'' (the improvement of the ductility of W and Mo at
the doping by Re). According to Ref. 11 the increase of the solubility of
interstitial impurities connected with the appearance of the precipitates of
the FK phases in particular W$_3$Re is the key feature in this phenomenon.
The results presented here allows us, on the one hand, to understand the
causes of the appearance of the precipitates and, on the other hand, to
point out the specific mechanism of their influence on the solubility of
interstitial impurities. Indeed, considerable dilatation in the cloud
surrounded the precipitate may lead to the essential energy gain at the
transfer of the interstitial atom from the host to the shell. It may prevent
the appearance of the carbide precipitates or impurity segregation at the
intergrain boundaries.

Note in the conclusion that the transition from the two-dimensional case
considered here to the three-dimensional one is not trivial since to
``dress'' the spherical precipitate by the cloud of compensated defects we
cannot restrict ourselves by the disclinations only and it is necessary to
introduce the density of disclination loops as an additional variable.
However the results presented here seem to be sufficient to demonstrate the
instability of the homogeneous state under some conditions.

To resume it is worthwhile to note that the present work show the
interconnection of some specific facts known in the material science with
non-trivial properties of three-dimensional space and therefore demonstrate
the general physical meaning of these facts.

This work is supported by Russian Basic Research Foundation, grants
95-02-05656 and 95-02-06426. We are grateful to A.M.Kosevich for helpful
discussions.

\newpage

\begin{figure}
\caption
{ The dependence of the
equilibrium values of precipitate radius $R$, disclination density
$\theta_1$ (in units of $\widetilde\theta _0$)
and the value of $\psi $ in Eq.(11) on the $\kappa $ parameter
(14).
}
\label{fig:1}
\end{figure}

\begin{figure}
\caption
{ The distribution of the dilatation $\varepsilon _{kk}$
in the percipitate (solid line) on the polar coordinate $r$ (R=4,
$\Delta =1$), and the dependence of {\it average} dilatation
$\left\langle \varepsilon_{kk}\right\rangle $ on the precipitate
radius $R$ (dashed line) in the units $\theta_1 (1-2\nu )/(2\pi (1-\nu ))$.
}
\label{fig:2}
\end{figure}

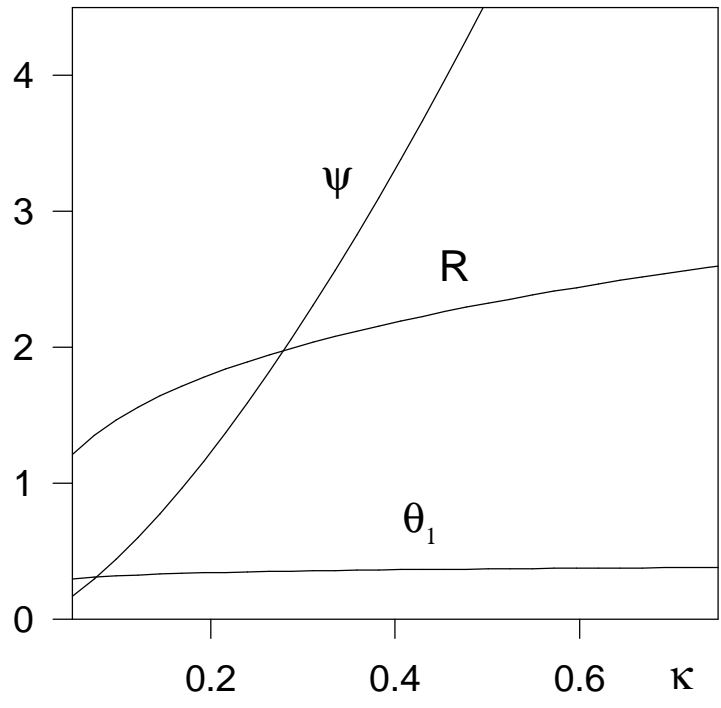
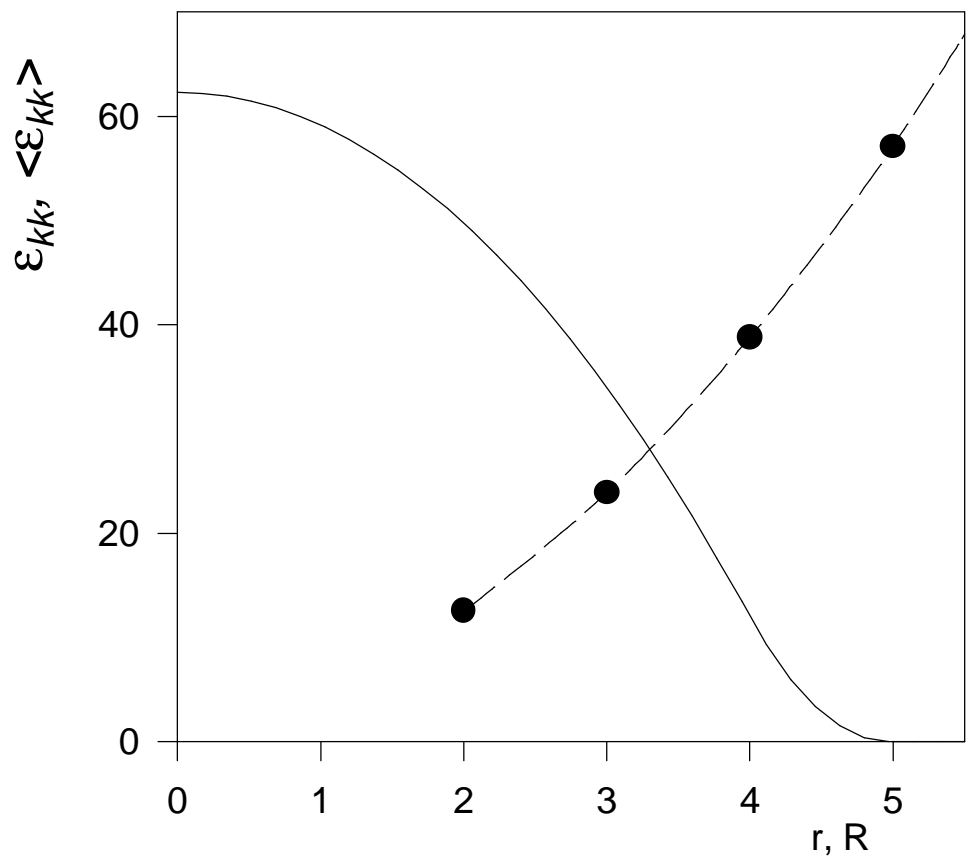
\end{document}